\documentclass{iauc}
\usepackage{graphicx}
\usepackage{amsmath}
\usepackage{upmath}

\title %% give here short title %% 
{The stellar populations of local dwarfs}

\author[Enrico V. Held]   %% give here short author list %%
{Enrico V. Held$^{1}$}
\affiliation{
$^1$Osservatorio Astronomico di
Padova, INAF, vicolo dell'Osservatorio 5, I-35122 Padova, Italy
}
\pubyear{2005}
\volume{198} %% insert here IAU Symposium No.
\pagerange{1--1}
\jname{Near-Field Cosmology with Dwarf Elliptical Galaxies}
\editors{B. Binggeli, H. Jerjen, eds.}
\begin{document}

\maketitle
\begin{abstract}
Recent progress in our knowledge of stellar populations in local dwarf
spheroidal galaxies is briefly discussed.  A few results are summarized
including wide field observations of stellar populations and their
spatial variations, studies of AGB and variable stars, extension to
near-infrared wavelengths, and the interpretation effort based on
synthetic color-magnitude diagrams and chemical evolution models.

%% add here a maximum of 10 keywords, to be taken form the file <Keywords.txt>.
\end{abstract}


%------------------------------------------------------------------
\begin{figure}
\begin{center}
\includegraphics[width=1.1\textwidth]{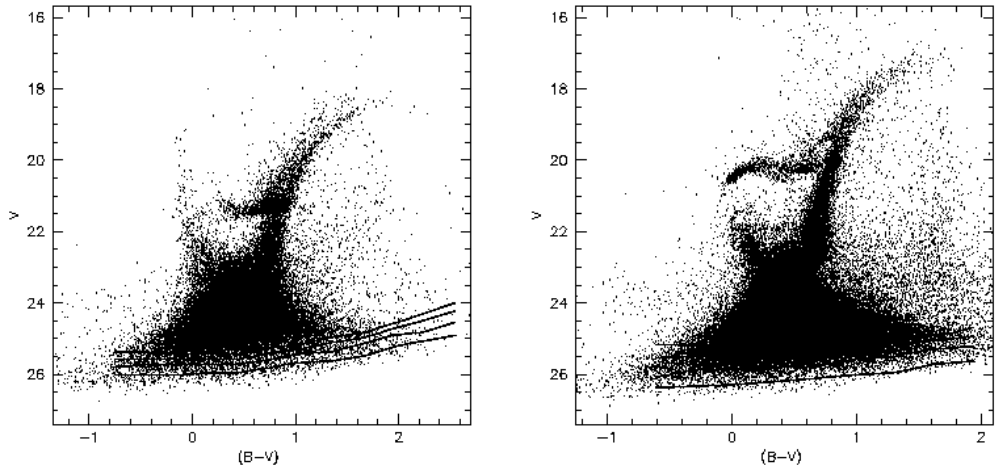}
\caption{Wide-field color-magnitude diagrams of Fornax ({\it left}) and
Sculptor ({\it right}) dwarf spheroidals based on ESO/WFI observations
(Rizzi et al. 2005, in prep.).
%% Iso-completeness contours obtained by simulating the photometry process
%% for artificial stars of known magnitudes are superimposed, indicating
%% the 10\% to 40\% levels of photometric completeness.
}
\label{evheld-fig1}
\end{center}
\end{figure}

%----------------------------------------------------------------------
\section{Introduction}

%% Dwarf spheroidal galaxies can be resolved into stars, and their stellar
%% content and evolution studied with the techniques of stellar astronomy.
%% From their color-magnitude diagrams, we can reconstruct (with some
%% degree of uncertainty) their star formation histories. With the advent
%% of 8--10m class telescopes, also the path of chemical enrichment in
%% their stars, and even the detailed abundance patterns of different
%% elements, become within reach of our instruments. Photometric and
%% spectroscopic information together give us the constraints we need to
%% fully model the evolution of dwarf galaxies. Thus observational methods
%% stellar astronomy can help us to understand the evolution of early type
%% dwarf galaxies on a cosmological scale.

The study of stellar populations of nearby dwarf galaxies provides a
powerful means of learning, albeit indirectly, about the evolution of
low mass galaxies at large look-back times.  Galaxies in the Local Group
are close enough that they can be resolved in the constituent stars and
studied using the classical tools of stellar population work, the HR
diagram in the first place. Diagnostics derived from the color-magnitude
diagram of local dwarf galaxies are used to pinpoint and, to some
extent, quantify the presence of young, intermediate-age (1-9 Gyr), and
old ($\gtrsim 10$ Gyr) stellar populations and to reconstruct their histories
of star formation. Recent advances with spectroscopy of individual stars
and nebulae provide complementary information to understand their
chemical enrichment histories.
I will focus here on the resolved stellar populations of dwarf
spheroidal (dSph) galaxies, with the exclusion of nearest Milky Way
companions and the dwarf galaxies around M\,31 and in nearby groups,
which are the subject of other contributions to this conference.
I refer the reader to existing reviews (Mateo \cite{mate98}, Da Costa
\cite{daco98}, van den Bergh \cite{vdb00}, Grebel \cite{greb00}, Tolstoy
\cite{tols03}, Gallagher \& Grebel \cite{jgal+greb04}) for earlier work
and a comprehensive account of stellar populations in Local Group dwarfs
as emerged from the last decade of HST and ground-based studies.

%------------------------------------------------------------
\section{The stellar populations of dwarf spheroidal galaxies}

\subsection{Stellar populations: recent results}

%% The observation of radial gradients (Harbeck et al. (\cite{harb+01}) and
%% asymmetric distributions (e.g., Monelli et al. \cite{mone+03}) in
%% stellar populations of different ages, along with the progress in
%% modeling star formation histories, makes it conceivable to think of a
%% spatially resolved approach to star formation and chemical enrichment
%% histories of nearby dwarf galaxies. However, a comprehensive approach
%% using both photometric and spectroscopic constraints is necessary to
%% model the evolution of dwarf spheroidals.

Direct age measurements of the main-sequence turnoff in nearby dSph, as
well as indirect evidence from horizontal branch stars in more distant
dwarfs, have shown that all dwarf spheroidals in the Local Group are
characterized by a common early epoch of star formation approximately
coeval to the Galactic globular clusters (10--13 Gyr ago) (e.g., Held et
al. \cite{held+00}; Grebel \cite{greb00}). This initial star formation
episode is predominant in some dSph, but the general evidence is
for multiple episodes or a continuous star formation
until very recent times.

Two examples of the color-magnitude diagrams (CMDs) of  ``young'' and
``old'' dwarf spheroidal galaxies are shown in Fig.~\ref{evheld-fig1}: in
both cases, the stellar populations show a complexity typical of a range
of age and metallicity.
Fornax (see Fig.~\ref{evheld-fig1}, left panel; also Saviane et
al. \cite{savi+00}) is an example of a ``young'' dwarf spheroidal: its
CMD is characterized by a ``blue plume'' of main-sequence stars
indicating an extended period of star formation, until a few hundred Myr
ago. The very wide red giant branch clearly shows the presence of a mix
of old and intermediate-age stars with different metal abundances.
Pont et al. (\cite{pont+04}) have recently studied the star formation
and chemical enrichment in a central region of Fornax by combining VLT
photometry and stellar spectroscopy.  Their observations suggest an 
increase in star formation rate and metallicity in the last
few Gyr: the metallicity distribution of Fornax is centered at [Fe/H]$
=-0.9$ with a long tail of metal-poor stars extending to [Fe/H]$ =-2.0$
and a number of stars as metal-rich as [Fe/H]$ =-0.4$.

The Carina dwarf is another example of dSph galaxy with a prominent
intermediate-age population.  With two main bursts occurring 5 and
11 Gyr ago, and a small population of stars on a younger main sequence
(up to 0.6 Gyr ago), Carina is the best example of an episodic star
formation history (see Monelli et al.  \cite{mone+03}; and
refs. therein).  This galaxy also provides the most dramatic example of
an ``age-metallicity degeneracy'', i.e. the balance of competing and
superposed effects of a younger age and an increasing metallicity on the
CMD distribution of red giant stars.
Yet its very narrow red giant branch, seemingly contrasting the
multiple star formation episodes, does not imply a small range in
[Fe/H]. A quantitative analysis of the CMD of Carina based on synthetic
diagrams shows that the narrowness of the RGB is indeed consistent with
the combined effects of the star formation and chemical enrichment
history (Rizzi et al. \cite{rizz+03}). The presence of a wide range in
metallicity is confirmed by spectroscopy (Koch et al. \cite{koch+04}).

%% The Carina dSph shows a strong gradient in the population of
%% intermediate-age He-burning stars, which are concentrated in the inner
%% regions.  Accordingly, the spatial variations in the SFH can be
%% reconstructed (Rizzi et al. \cite{rizzi-sfh}), with the intermediate-age
%% episode of star formation becoming increasingly prominent toward the
%% center.

%----------------------------------------
%\subsection{Stellar population gradients}

Figure~\ref{evheld-fig1} (right panel) shows the color-magnitude diagram
of Sculptor.  Sculptor is an ``old'' dwarf spheroidal galaxy lacking a
significant star formation in the last 5 Gyr.  It shows a particularly
strong radial variation in the HB morphology index, which is accompanied
by a spatial gradient in the RGB stars in the outer regions (Harbeck et
al. \cite{harb+01}; Tolstoy et al. \cite{tols+04}; Babusiaux et
al. \cite{babu+05}). The composite nature of the RGB (in particular the
existence of a substantial metal-poor component) suggested by previous
studies is not confirmed by recent photometry 
(Babusiaux et al. \cite{babu+05}).
The presence of these gradients in both the HB morphology and the RGB
implies that a significant metallicity gradient must be present (Rizzi
et al. \cite{rizzi-sfh}).  Using spectroscopic estimates of [Fe/H] for
RGB stars, Tolstoy et al. (\cite{tols+04}) indeed find two components
among red giant stars, with the metal-rich component more centrally
concentrated than the metal-poor stars.

Similar CMD morphology gradients are seen in Sextans dSph (Harbeck et
al. \cite{harb+01}; Lee et al.  \cite{mglee+03}).  Our own synthetic CMD
analysis suggests that in this case the stellar population gradient is
driven by both metallicity and age variations.  A hint for the origin of
the changes in the stellar populations within Sextans dSph may also come
from a comparison with internal kinematics. Indeed, Kleyna et
al. (\cite{kley+04}) associate a cold, inner stellar component having
velocity dispersion close to zero with the change in the stellar
populations near the center.

Other recent studies of old, metal-poor dwarf spheroidals include Draco,
Ursa Minor, and Cetus (Aparicio et al. \cite{apar+01}; Carrera et
al. \cite{carr+02}; Wyse et al. \cite{wyse+02}; Sarajedini et
al. \cite{sara+02}). In Draco and Ursa Minor, the metallicity
distribution is narrow yet with a measurable spread and slightly
different shapes indicating different star formation histories
(Bellazzini et al. \cite{bell+02}).

%-----------------------------------------------------------------
\subsection{An extended star formation in old dwarf spheroidals ?}

One interesting result of wide-field observations is that star formation
may have continued at a low rate up to a recent (a few Gyr) epoch even in
predominantly old dwarf spheroidals.
In the CMD of Sextans, Ursa Minor, and Sculptor (see
Fig.~\ref{evheld-fig1}), a ``plume'' of stars is obviously present above
the main-sequence turnoff.  If those stars are main sequence stars,
their presence implies a modest rate of star formation up to $\sim 2$ Gyr
ago, after the bulk of stars was formed at old epochs. However, the
possibility cannot be ruled out that these stars are ``blue stragglers''
like those found in globular clusters, which are likely to be the
evolved products of mass transfer in old binary systems (see Lee et
al. \cite{mglee+03}).
The recent studies of Ursa Minor and Sextans conclude that the ``blue
plume'' is more likely to be made of old GC-like ``blue stragglers''
rather than intermediate-age normal main-sequence stars (Carrera et
al. \cite{carr+02}; Lee et al. \cite{mglee+03}).  A similar
interpretation is proposed for the ``blue plume'' stars in Sculptor
(Rizzi et al. \cite{rizzi-sfh}), while a different conclusion is reached
by Aparicio et al. (\cite{apar+01}) for Draco, where an analysis of the
red clump and subgiant branch stars supports the existence of an
intermediate age population, up to 2--3 Gyr ago.
These different interpretations obviously affect our picture of early
star formation in old dwarf spheroidals and their relation to the
reionization era (see Grebel \& Gallagher \cite{greb+jgal04}).

%% The luminosity function of low-mass main sequence stars in dwarf
%% spheroidals keeps track of the stellar IMF in dwarf galaxies at old
%% epochs, an issue of great importance to interpret observations of high
%% redshift objects. Ursa Minor formed most of its stars more than 12 Gyr
%% ago, i.e. at redshift $z \gtrsim 2$, and it is close enough (only 70
%% kpc) that its low-mass stellar IMF can be derived from star counts (Wyse
%% et al. \cite{wyse+02}). It turns out to be indistiguishable from those
%% of globular clusters in the halo of the Milky Way, indicating that the
%% low-mass IMF is invariant over most parameters (metallicity, age,
%% star-formation rate, dark matter content, etc ...).

%------------------------------------------------------------
\subsection{Evolved stellar populations}

The intermediate-mass stars on the extended asymptotic giant branch
(AGB), in particular the carbon-rich (C) stars, are important tracers of
intermediate-age populations in dwarf galaxies (for a review of AGB
stars in Local Group galaxies, see Groenewegen \cite{groe04}).
Because of their high luminosities and distinctive spectral features, C
stars are relatively easy to identify on the basis of narrow-band and 
near-infrared colors. Carbon star surveys have been conducted by
different groups (e.g., Kerschbaum et al. \cite{kers+04}; Battinelli \&
Demers \cite{batt+deme05}; Harbeck et al.  \cite{harb+04}).
The formation of C stars is both a function of the star formation
history and metallicity.  Because of this dependence on metallicity, not
only is the ratio between carbon-rich and oxygen-rich cool giants, C/M,
globally a function of galaxy metallicity (see, e.g., Battinelli \&
Demers \cite{batt+deme05}, and references therein) but also it can be
used to map {\it local} abundance variations across galaxies.  New
models including revised effects of variable opacity in carbon-rich
stars are now able to better reproduce the observed distribution of
luminous AGB stars in near-infrared CMDs (Marigo \cite{mari02}).
Since even a few luminous AGB stars can significantly contribute to the
integrated light of galaxies, studying the behavior of AGB stars in
nearby dwarfs represents an important step towards a successful
intepretation of the spectra of distant dwarf galaxies.  Evolutionary
population synthesis models focusing on intermediate-age populations and
their contribution to the integrated light have been presented by
Mouhcine \& Lan\c{c}on (\cite{mouh+lanc03}).

%------------------------------------------------------------
%\subsection{Evolved stellar population: near-infrared observations}

The cool AGB as well as the RGB stars are best observed in the
near-infrared, where bolometric corrections to intrinsic luminosities
and effective temperatures are smaller, making interpretation of the
observations on the basis of stellar evolution models more reliable.
The optical-near infrared colors of RGB stars are more sensitive to
metallicity than optical (or infrared) colors alone, in particular for
old, metal-poor systems. This alleviates the age-metallicity degeneracy
and allows an improved estimate of both mean metallicities and
metallicity distributions (Babusiaux et al. \cite{babu+05}; Gullieuszik
et al., this conference).
If we exclude the Sagittarius dSph, which is close enough to be studied
with the 2MASS database (e.g., Cole \cite{acol01}), the sensitivity and
small size of near-infrared instrumentation have limited the number of
studies to very few dwarf spheroidals (Menzies et al. \cite{menz+02};
Pietrzy{\' n}ski et al. \cite{piet+03}; Babusiaux et
al. \cite{babu+05}). Ongoing wide-area studies of dSph galaxies by our
group as well as other teams aim at filling this gap, especially using
the new wide-field instruments.

%------------------------------------------------------------
\subsection{Variable stars as stellar population tracers}

Pulsating variable stars have been studied in dSph galaxies for many
years (Mateo \cite{mate98}).  Besides providing independent distance
estimates to the galaxies, variable stars can trace the presence and
properties of different stellar generations.  In particular, RR\,Lyrae
variables originate from the oldest stars ($\gtrsim 10$ Gyr), 
so that an
old stellar population can be detected even when the data are too
shallow or confusion-limited to reach the main sequence turnoff of
low-mass stars. Measurements of RR Lyrae metallicity based on their
pulsational properties (period and amplitude of the light curve) help 
us to reconstruct the metal enrichment history of dwarf spheroidals.
In addition, the radial distribution of variable stars in different
classes can be used to map the gradients in their parent stellar
populations (e.g., Gallart et al. \cite{gall+04}).

Several recent studies have
addressed the properties of variable stars in dSph galaxies (e.g.,
Leo\,I: Held et al. \cite{held+01}; 
Fornax: Bersier \& Wood \cite{bers+wood02}, Poretti et al. 2005, in prep.;
Carina: Dall'Ora et al. \cite{dall+03};
Draco: Bonanos et al.  \cite{bona+04}; 
Phoenix: Gallart et al. \cite{gall+04}; 
and references therein).
The RR\,Lyrae in most dSph galaxies appear to be intermediate between
those in Oosterhoff type~I and type~II globular clusters (e.g., Dall'Ora
et al. \cite{dall+03}), a characteristic shared by the globular clusters
belonging to Fornax dSph (Mackey \& Gilmore \cite{mack+gilm03}).

Other classes of variable stars bear information on the stellar content
of local dwarfs.  An interesting class is Anomalous Cepheids, which are
found numerous in dSph galaxies, where their number per unit luminosity
seems to be correlated to the galaxy metallicity and luminosity 
(e.g., Pritzl et al. \cite{prit+05}).
That Anomalous Cepheids are related to the short-period Classical
Cepheids found in (dwarf) irregular galaxies is still a matter of debate
(Dolphin \cite{dolp02}; Baldacci et al. \cite{bald+04}; Marconi et
al. \cite{mmar+04}; Gallart et al. \cite{gall+04}; Pritzl et
al. \cite{prit+05}).
%
%% Recent models interpret their pulsational properties and 
%% location in the CMD as intermediate-mass (1--2 M$_\odot$) stars crossing
%% the instability strip during their core He-burning phase (see, e.g.,
%% Marconi et al. \cite{mmar+04}).  
%
As such, the Anomalous Cepheid would trace the intermediate-age
populations rather than the old component.

%------------------------------------------------------------
\subsection{Modeling the star formation histories}

Presently, the main challenge is to interpret observational facts in
terms of galaxy evolution and  understand the chemical enrichment and
star formation histories of dSph galaxies (see, e.g., Gallagher \&
Grebel \cite{jgal+greb04}).
Disentangling the age and metallicity variations of the stellar
populations is a complex task that models based on evolutionary tracks
and synthetic color-magnitude diagrams have recently undertaken
(Hernandez et al. \cite{hern+00}; Ikuta \& Arimoto \cite{ikut+arim02};
Dolphin \cite{dolp02}; Rizzi et al. \cite{rizzi-sfh}; Pont et
al. \cite{pont+04}). While in many cases the models assume a metal
enrichment law, 
some chemical evolution models have started to shed some on the
interplay of star formation and chemical evolution of dwarf spheroidal
galaxies (e.g., Ikuta \& Arimoto \cite{ikut+arim02}; Carigi et al.
\cite{cari+02}; Lanfranchi \& Matteucci \cite{lanf+matt04}).  For
example, in the Lanfranchi \& Matteucci (\cite{lanf+matt04}) models, the
metallicity distributions of stars in dSph galaxies are predicted using
the star formation histories derived from the CMDs together with
physical constraints such as the mass and gas content, and the observed
abundance patterns from high-resolution spectroscopy (e.g., Tolstoy et
al. \cite{tols+03}; Shetrone \cite{shet04}).
The overall properties of dwarf spheroidals imply very low star
formation rates and a high efficiency of galactic winds in order to
reproduce the observed gas fraction and abundance ratios, a conclusion
on which the different models mostly agree.

Environmental factors, such as interaction with the Galaxy and M\,31,
may also influence the evolution of dSph galaxies.
In the case of the Milky Way satellites, our understanding of the
relationship between history of star formation and {\it orbital motions}
can benefit from recent analyses of proper motions.  For Fornax, the
work of Dinescu et al. (\cite{dine+04}) suggests that the termination of
star formation about 200 Myr ago indicated by color-magnitude diagrams
coincides with the time when Fornax crossed the Magellanic Plane.
However, in the case of Carina, the history of star formation is not
explained by passages through perigalacticon and the Galactic disk --
the interval between multiple passages being too short to explain the
main star formation episode about 6-7 Gyr ago (Piatek et
al. \cite{piat+03}).  Thus, while interaction with the Milky Way is
suggested to be able to trigger or stop star formation (by gas removal),
it is certainly not the only driving factor of their star formation
histories.

%------------------------------------------------------------
%\section{Acknowledgements}
\bigskip

I wish to thank Ivo Saviane, Yazan Momany, Luca Rizzi, Marco
Gullieuszik, and Gianpaolo Bertelli, my collaborators on stellar
population research, and the organizers of this enjoyable meeting.  The
search for variable stars in local dwarfs is done in collaboration with
teams led by G. Clementini, E. Poretti, H. Smith, and M. Catelan.

%-----------------------------------------------------------------

\end{document}